\input psfig.sty
\newif\iftitlepage
\titlepagetrue
\newtoks\titlepagehead  \titlepagehead={\hfil}
\newtoks\titlepagefoot  \titlepagefoot={\hfil}
\newtoks\runningauthor  \runningauthor={\hfil}
\newtoks\runningtitle   \runningtitle={\hfil}
\newtoks\evenpagehead   \newtoks\oddpagehead
\evenpagehead={\hfil\the\runningauthor\hfil}
\oddpagehead={\hfil\the\runnintitle\hfil}
\newtoks\evenpagefoot \evenpagefoot={\hfil\tenrm\folio\hfil}\newtoks\oddpagefoot 

\font\bf=cmbx12
\oddpagefoot={\hfil\tenrm\folio\hfil}\headline{\iftitlepage\the\titlepagehead\else\ifodd\pageno\the\oddpagehead\else\the\evenpagehead\fi\fi}
\footline={\iftitlepage\the\titlepagefoot\global\titlepagefalse
\else\ifodd\pageno\the\oddpagefoot\else\the\evenpagefoot\fi\fi}
\newcount\choiceno
\newcount\probno\probno=-1
\def\prob{\futurelet\next\prob@}
\def\prob@{\ifx\next<\expandafter\prob@@\else\expandafter\prob@@@\fi}
\def\prob@@<#1>{\problem<#1>\everypar={\vglue-20pt\endproblem}}
\def\prob@@@{\problem\everypar={vg;ue-20pt\endproblem}}
\def\oaint{{\oint\limits_{C^\prime}\!\!\!\!\;{}_{\hat{}}}}

\def\bra{{\rm <}}                  
\def\ket{{\rm >}}                  
\newcount\q \q=0
\newcount\qq  \qq=0
\def\nref {\global\advance\q by1\item{\bf\the\q.}}
\def\nrf {\global\advance\qq by1 \eqno{(\the\qq)}}
\magnification=\magstep1
\baselineskip 18 true pt    \parskip=0pt plus 5pt
\parindent 0.25in
\hsize 6.25 true in \hoffset .20 true in
\vsize 8.5 true in \voffset .1  true in
\evenpagehead={\vbox{\line{\tenrm\the\runningtitle\hfil\tenbf Page\ \folio}
\medskip\hrule\bigskip}}
\oddpagehead={\vbox{\line{\tenrm\the\runningtitle\hfil\tenbf Page\ \folio}
\medskip\hrule}}
\evenpagefoot={\hfil} \oddpagefoot={\hfil}
\runningauthor={R. L. Hall {\it et al}}
\runningtitle={\it Matrix Elements for $\dots$}
\centerline{\bf Matrix Elements for a Generalized  }
\centerline{\bf Spiked Harmonic Oscillator}
\medskip
\vskip 0.25 true in
\centerline{Richard L. Hall, Nasser Saad
\footnote{$^\dagger$}{Present address: Department of mathematics, faculty of science, 
Notre Dame University, Beirut, Lebanon}}
\centerline{and}
\centerline{Attila B. von Keviczky}
\bigskip
{\leftskip=0pt plus 1fil
\rightskip=0pt plus 1fil\parfillskip=0pt\baselineskip 18 true pt
\obeylines
Department of Mathematics and Statistics,
Concordia University,
1455 de Maisonneuve Boulevard West,
Montr\'eal, Qu\'ebec,
Canada H3G 1M8.\par}
\bigskip
\vskip 0.5 true in
\centerline{\bf Abstract}
\noindent Closed form expressions for the singular-potential integrals 
$\bra m|x^{-\alpha}|n\ket$ are obtained with respect to the Gol'dman and Krivchenkov eigenfunctions for the singular potential $Bx^2+{A\over x^2}$, $B>0, A\geq 0$. The formulas obtained are generalizations of those found earlier by use of the odd solutions of the Schr\"odinger equation with the harmonic oscillator potential [Aguilera-Navarro et al, J. Math. Phys. 31, 99 (1990)].
\vfill\eject
\noindent{\bf I. Introduction}
\medskip
In 1990 Aguilera-Navarro {\it et al}$^1$ employed a perturbative scheme to 
provide a variational analysis for the lowest eigenvalue of the Schr\"odinger 
operator
$$H=-{d^2\over {dx^2}}+x^2+{\lambda\over x^\alpha},\quad 0\leq x<\infty,$$
where $\alpha$ is a positive constant. Writing $H\equiv H_0+\lambda V$ with 
$H_0$ standing for the harmonic oscillator Hamiltonian and $V=x^{-\alpha}$, 
Aguilera-Navarro {\it et al}$^1$ used the basis set
$$\psi_n(x)=A_ne^{-x^2/2}H_{2n+1}(x),\quad A_n^{-2}= 2^{2n+1} (2n+1)!,\quad 
n=0,1,2,\dots\nrf$$
constructed from the normalized solutions of $H_0\psi=E\psi$ to evaluate the matrix 
elements of $H$. They found that
$$H_{m+1,n+1}\equiv (3+4n)\delta_{m,n}+\lambda \bra m|x^{-\alpha}|n\ket\quad 
m,n=0,1,2,\dots,N-1,\nrf$$
where
$$\eqalign{\bra m|x^{-\alpha}|n\ket&=(-1)^{n+m}{{\sqrt{(2n+1)!(2m+1)!}}\over 
{2^{n+m}n!m!}}\cr
&\times \sum\limits_{k=0}^m (-1)^k {m\choose k}{{\Gamma(k+{{3-\alpha}\over 
2})\Gamma(n-k+{\alpha\over 2})}\over {\Gamma(k+{3\over 2})\Gamma(-k+{\alpha\over 
2})}}.\cr}\nrf$$
The aim of this article is to extend these results to treat the more general
spiked harmonic oscillator Hamiltonian
$$H=-{d^2\over {dx^2}}+Bx^2+{A\over x^2}+{\lambda\over x^{\alpha}}\quad B>0, 
A\geq 0.\nrf$$
The particular case of $A=0$ and $B=1$ allows us, of course, to recover the results 
of Aguilera-Navarro {\it et al}$^1$ as a special case.  The existence of all the 
exact eigenfunctions for a problem with a singular potential is the principal 
motivation for the present work: we expect that such a set of functions will be 
more effective for the analysis of other singular problems than are the Hermite functions generated, as they are, by the non-singular harmonic-oscillator potential $x^{2}.$ 

The article is organized as follows. In Sec. II we provide an orthonormal set of 
functions that we use to compute the matrix elements  of the Hamiltonian (4). In 
Sec. III we prove that this set of functions is complete in the sense that it is 
a basis for $L^{2}(0,\infty).$  We then compute the matrix elements using this 
basis in Sec. IV. In Sec. V the comparison with the result of Aguilera-Navarro 
{\it et al}$^1$ are presented for the special case $B=1$ and $A=0$:
where we shall point out some errors in their value for the matrix element $\bra\psi_3|x^{-\alpha}|\psi_3\ket$ as quoted in the appendix of Ref. [1]. 
\bigskip
\noindent{\bf II. An orthonormal basis}

\noindent Gol'dman and Krivchenkov$^2$ have provided a clear description of the 
exact solutions of the following one dimensional Schr\"odinger equation (in 
units $\hbar=2m=1$)
$$-\psi^{\prime\prime}+V_0\bigg({a\over x}
-{x\over a}\bigg)^2\psi=E_n\psi,\qquad x\in[0,\infty)\nrf$$
with $\psi$ satisfying the Dirichlet boundary condition $\psi(0)=0$.
They showed that the energy spectrum in terms of the parameters $V_0$ and $a$ is 
given by
$$E_n={4\over a}\sqrt{V_0}\bigg\{n+{1\over 2}+{1\over 
4}\bigg(\sqrt{1+4V_0a^2}-2a\sqrt{V_0}\bigg)\bigg\}.\nrf$$
To simplify notation we introduce the parameters $B=V_0a^{-2}$ and $A=V_0a^2$, 
and obtain thereby an exact solution to the Schr\"odinger equation with the 
singular potential
$$
V(x)=B x^2+{A\over x^2}, \quad B>0, A\geq 0,\nrf
$$
where the energy spectrum is now given in terms of the parameters $A$ and $B$ by
$${E}_n=\sqrt{B}(4n+2+\sqrt{1+4A}),\quad n=0,1,2,\dots.\nrf$$
The wave functions have the form
$$\bra x|n\ket\equiv \psi_n(x)=C_nx^{{1\over 2}(1+\sqrt{1+4A})}e^{-{1\over 
2}\sqrt{B} x^2}{}_1F_1(-n,1+{1\over 2}\sqrt{1+4A};\sqrt{B}x^2),\nrf
$$
where $n=0,1,2,\dots$ and ${}_1F_1$ is the confluent hypergeometric function$^3$
$${}_1F_1(a,b;z)=\sum\limits_k {{(a)_kz^k}\over {(b)_kk!}}.$$
The Pochhammer symbols $(a)_k$ are defined as
$$(a)_k=a(a+1)(a+2)\dots(a+k-1)={{\Gamma(a+k)}\over {\Gamma(a)}},\quad 
k=1,2,\dots\nrf$$
where $\Gamma(a)$ is the gamma function. Note that we have corrected the misprint of Ref. [4] for the power of $x$ in the wavefunctions (9). The constant $C_n$ is determined from the normalization condition
$$\int_0^{\infty}\psi_n^2(x)dx=1,$$
which requires use of the identity$^4$
$$\eqalign{&\int\limits_0^\infty e^{-\lambda 
x}x^{\nu-1}[{}_1F_1(-n,\gamma;kx)]^2dx=
{{n!\Gamma(\nu)}\over {k^{\nu}\gamma(\gamma+1)\dots (\gamma+n-1)}} \cr
&\times\bigg\{1+\sum_{s=0}^{n-1}{{n(n-1)\dots(n-s)(\gamma-\nu-s-1)(\gamma-\nu-s)
\dots(\gamma-\nu+s)}\over 
{[(s+1)!]^2\gamma(\gamma+1)\dots(\gamma+s)}}\bigg\},\cr}\nrf
$$
thereby yielding
$$C_n^2=
{{2B^{{1\over 2}+{1\over 4}\sqrt{1+4A}}\Gamma(n+1+{1\over 2}\sqrt{1+4A})}
\over 
{{n![\Gamma(1+{1\over 2}\sqrt{1+4A}})]^2}}.\nrf
$$
In order to prove that the $\psi_n(x)$, $n=0,1,2,\dots,$ defined by (9) are 
orthonormal, we have to demonstrate
$$\int\limits_0^\infty\psi_n(x)\psi_m(x)dx=0\quad (n\neq m).\nrf$$
We know however, that $\psi_n(x)$ and $\psi_m(x)$ satisfy the Schr\"odinger 
equations
$$\psi_n^{\prime\prime}+(E_n-Bx^2-{A\over x^2})\psi_n=0,\qquad 
E_n=\sqrt{B}(4n+2+\sqrt{1+4A}),\nrf$$
$$
\psi_m^{\prime\prime}+(E_m-Bx^2-{A\over x^2})\psi_m=0,\qquad 
E_m=\sqrt{B}(4m+2+\sqrt{1+4A}).\nrf$$
Multiplying (14) by $\psi_m$ and (15) by $\psi_n$ and then subtracting the 
resulting equations, we obtain 
$${d\over {dx}}(\psi_m\psi_n^\prime-\psi_n\psi_m^\prime)+\sqrt{B}(n-m)\psi_n\psi_m=0.$$
After integrating this equation over $[0,\infty)$ and using 
$\psi_n(0)=\psi_m(0)=0$ and $\psi_n(\infty)=\psi_m(\infty)=0$, we get 
$$(n-m)\int\limits_0^\infty\psi_n(x)\psi_m(x)dx=0,$$
which proves (13). Thus we obtain the following identity

$$
\int\limits_0^\infty e^{-\lambda x^2} x^{2\gamma-1}{}_1F_1(-n,\gamma;\lambda 
x^2){}_1F_1(-m,\gamma;\lambda x^2)dx=\left\{\eqalign{&0\qquad\qquad\quad\  {\sl 
if}\quad n\neq m,\cr
&{1\over 2}{{n!\Gamma(\gamma)}\over {\lambda^\gamma(\gamma)_n}}\qquad {\sl 
if}\quad n= m,\cr}\right.
$$
wherein the confluent hypergeometric functions ${}_1F_1$ are defined as follows$^3$: 
$$
{}_1F_1(-n,\gamma;r)\equiv -{1\over {2\pi i}}{{\Gamma(n+1)\Gamma(\gamma)}\over 
\Gamma(n+\gamma)}
\oaint\ e^{tr}(-t)^{-n-1}(1-t)^{\gamma-n-1}dt
=$$
$$
{{\Gamma(\gamma)}\over 
{\Gamma(\gamma+n)}}r^{1-\gamma}e^rD^n(r^{\gamma+n-1}e^{-r})
={{\Gamma(\gamma)}\over {\Gamma(\gamma+n)}}r^{1-\gamma}(D-1)^n(r^{\gamma+n-1})
\nrf
$$
for any simply closed rectifiable contour $C^\prime$ starting at $1$ and 
enclosing the straight line segment from $0$ to $1$ in the complex plane, as 
illustrated in Fig. 1.
With $\gamma=1+{1\over 2}\sqrt{1+4A}$ and $\lambda=\sqrt{B}$,
the set of $L^2(0,\infty)$-functions
$$\psi_n(x)=C_n x^{\gamma-{1\over 2}}e^{-{1\over 2}\lambda 
x^2}{}_1F_1(-n,\gamma;\lambda x^2),\quad n=0,1,2,\dots\nrf
$$
constitutes a orthonormal system of the Hilbert space $L^2(0,\infty)$.
\bigskip
\noindent{\bf III. Proof of Completeness}

For the orthonormal functions $\{\psi_{n}\}$ to qualify as a {\it basis} for 
$L^2(0,\infty)$, we must demonstrate the density of the linear manifold 
generated by these functions in the topology induced by the norm determined by 
the inner product
$\bra\cdot\ |\ \cdot\ket$. This is equivalent to showing that if $<\psi_n|f>=0$ 
for all $n=0,1,2,\dots$, then $f=0$ a.e. on $(0,\infty)$. To this end we note 
that out of the fourth expression of (16) follows 
$${}_1F_1(-n,\gamma;\lambda x^2)=\sum\limits_{k=0}^n {n\choose k} 
{{(-\lambda)^{n-k}\Gamma(\gamma)}\over {\Gamma(\gamma+n-k)}} x^{2(n-k)}.\nrf
$$
Hence the basis representation of the functions (vectors) 
$$\{{}_1F_1(-n,\gamma;\lambda x^2),
{}_1F_1(-(n-1),\gamma;\lambda x^2),\dots, {}_1F_1(-1,\gamma;\lambda 
x^2),{}_1F_1(-0,\gamma;\lambda x^2)\}$$ 
in terms of the basis 
$$
\{x^{2n},x^{2(n-1)},\dots,x^2,1\}
$$ 
is achieved by a lower triangular $(n+1)\times (n+1)$ matrix, whose diagonal 
entries are 
${{(-\lambda)^{n-k}\Gamma(\gamma)}\over {\Gamma(\gamma+n-k)}}$ for 
$k=0,1,2,\dots,n$ - i.e. this matrix is invertible provided $\lambda\neq 0$. 
Thus each  $x^{2n}$ is a unique linear combination of the $n+1$ functions 
${}_1F_1(-(n-k),\gamma;\lambda x^2)$ for $k=0,1,2,\dots,n$, which conclusion 
carries over to the $2n$-th degree Taylor polynomial 
$$e_n(-{\mu x^2\over 2})=\sum\limits_{k=0}^n {1\over k!}\bigg(-{\mu x^2\over 
2}\bigg)^k$$
of $e^{-{\mu x^2\over 2}}$ about the point $0$, where $\mu$ is an arbitrary 
parameter.

Let $f$ be an $L^2(0,\infty)$-function orthogonal to each of the $\psi_n$, 
which is equivalent to saying
$$ <e_n(-{\mu\ {\cdot\ }^2\over 2}\ )|f>=\int\limits_0^\infty x^{\gamma-{1\over 
2}}e_n(-{\mu x^2\over 2}){f(x)}dx=0\nrf$$
for all $n=0,1,2,\dots$. Here we note that 
$$x^{\gamma-{1\over 2}}e^{-{\lambda x^2\over 4}}{f(x)}$$
in an $L^1(0,\infty)$-function whose absolute value majorizes 
$x^{\gamma-{1\over 2}}e^{-{\lambda x^2\over 4}}e^{{|\mu | x^2\over 4}}{f(x)}$ for 
$|\mu|\leq {\lambda\over 4}$ and consequently also $x^{\gamma-{1\over 
2}}e^{-{\lambda x^2\over 4}}
e_n({-{\mu x^2/4}}){f(x)}$. 

\noindent Because $x^{\gamma-{1\over 2}}e^{-{\lambda x^2\over 4}}
e_n({-{\mu x^2/4}}){f(x)}$ converges to $x^{\gamma-{1\over 
2}}e^{-{\lambda x^2\over 2}}
{f(x)}$ a.e. on $(0,\infty)$ as $n\rightarrow \infty$, we conclude by 
means of the Lebesgue dominated convergence theorem$^5$ that we may replace $e_n(-{\mu x^2/2})$ by $e^{-{\mu x^2\over 2}}$ in Eq.(19) for all complex numbers $\mu$ such that $|\mu|\leq {\lambda\over 4}$, which, after setting $x=\sqrt{2t}$, yields the Laplace-Transform expression
$$
{\bf \cal L}\{F\}(z)=\int\limits_0^\infty e^{-zt}(\sqrt{2t})^{\gamma-{3\over 2}} f(\sqrt{2t})dt=0,\quad |z-\lambda|\leq {\lambda\over 4}.\nrf$$
However, the Laplace transform of the measurable Laplace-transformable function 
$F(t) =e^{-zt}(\sqrt{2t})^{\gamma-{3\over 2}} f(\sqrt{2t})$ 
defines a holomorphic function of variable $z$ in the right half plane $\Re (z)>0$ 
vanishing in the disc $|z-\lambda|\leq {\lambda\over 4}$. By uniqueness of the 
analytic function$^6$ the Laplace transform of the function must vanish in the 
right half plane, specifically ${\bf \cal L}\{F\}(s)=0$ for all $s$ on the 
interval $(0,\infty)$. Further, the Laplace transform determines $F(t)$ uniquely$^7$ 
a.e. in $t$ on $(0,\infty)$, hence $F(t)=0$ a.e. in $t$ or $f$ is the zero 
$L^2(0,\infty)$-function. Consequently, $\{\psi_n:n=0,1,2,\dots\}$ is an 
orthonormal basis of  $L^2(0,\infty)$.
\bigskip
\noindent{\bf IV. The matrix elements $\bra m|x^{-\alpha}|n\ket$}

Let us now split the Hamiltonian (5) into an $H_0$ part
$$H_0=-{d^2\over {dx^2}}+Bx^2+{A\over x^2},\quad x\geq 0\nrf$$
and a perturbation
$$H_I=\lambda/x^\alpha.$$ The eigenstates of $H_0$ are now given by (9) and 
their unperturbed energy is given by (8). All we need to do is to evaluate the 
matrix elements $<m|x^{-\alpha}|n>$ using the basis (9), namely
$$\eqalign{<m|x^{-\alpha}|n>=&C_nC_m\int\limits_0^{\infty} 
e^{-\sqrt{B}x^2}x^{-\alpha+1+\sqrt{1+4A}}
 {}_1F_1(-n,1+{1\over 2}\sqrt{1+4A};\sqrt{B}x^2)\cr&\times{}_1F_1(-m,1+{1\over 
2}\sqrt{1+4A};\sqrt{B}x^2)dx,\cr\alpha < 2 + \sqrt{1+4A}\cr}
\nrf$$
This is equivalent to
$$<m|x^{-\alpha}|n>={{C_nC_m}\over 2}B^{-{1\over 
2}(-\alpha+2+\sqrt{1+4A})}\times I,\nrf$$
where
$$
I=\int\limits_0^{\infty} 
e^{-r}r^{\gamma-s}{}_1F_1(-n,\gamma;r){}_1F_1(-m,\gamma;r)dr\nrf
$$
with $r=\sqrt{B}x^2, \gamma=1+{1\over 2}\sqrt{1+4A}$, and $s=1+{\alpha\over 2}$. 

>From the Fubini-Tonneli theorem$^5$ combined with the Leibniz formula for 
differentiating
the product of two functions, as well as exponential shift from the third 
expression to the fourth in Eq.(16), with $n$ replaced by $m$, we find that $I$ 
is given by the expression
$$\eqalign{I&=(-1)^n{{n![\Gamma(\gamma)]^2}\over 
{\Gamma(n+\gamma)\Gamma(m+\gamma)}}({2\pi i})^{-1}\oaint\ 
t^{-n-1}(1-t)^{\gamma-n-1}\int\limits_0^\infty e^{-(1-t)r}r^{1-s}\cr
&\times \bigg[\sum\limits_{k=0}^m (-1)^k {m \choose 
k}(\gamma+m-1)(\gamma+m-2)\dots (\gamma+k)r^{\gamma+m-1-(m-k)}\bigg]drdt.\cr}
$$
Further, owing to the fact that the simply closed rectifiable contour $C^\prime$ 
lies to the left of the complex number 1, Fig. 1, we have
$$\int\limits_0^\infty 
e^{-(1-t)r}r^{\gamma-s+k}dr=\Gamma(\gamma-s+k+1)(1-t)^{-\gamma+s-k-1}
$$
and our expression for $I$ thereby reduces to
$$\eqalign{I=&(-1)^n{{n![\Gamma(\gamma)]^2}\over 
{\Gamma(n+\gamma)\Gamma(m+\gamma)}}\sum\limits_{k=0}^m (-1)^k {m \choose 
k}{{\Gamma(m+\gamma)\Gamma(\gamma-s+k+1)}\over {\Gamma(k+\gamma)}}\cr
&\times ({2\pi i})^{-1}\oaint\ t^{-n-1}(1-t)^{s+n-k-2}dt.\cr}\nrf
$$
Since the contour $C^\prime$ has $0$ in its inside, Fig. 1, and the integrand has a weak singularity at $1$ (in consequence of $\Re (s+n-k-2)>-1$), the Cauchy integral 
formula lets us write the contour integral multiplied by $(2\pi i)^{-1}$ as the $n$-th derivatives of  $(1-t)^{s+n-k-2}$ evaluated at $t=0$. Utilizing thereafter the Pochhammer symbol in Gamma function format Eq.(10), we arrive at
$$I={{[\Gamma(\gamma)]^2}\over {\Gamma(n+\gamma)}}
\sum\limits_{k=0}^m (-1)^k {m\choose k}{{\Gamma(\gamma-s+k+1)\Gamma(s+n-k-1)}\over 
{\Gamma(\gamma+k)\Gamma(s-k-1)}}.\nrf$$
Therefore, the matrix elements are given by
$$\eqalign{\bra m|x^{-\alpha}|n\ket=&
(-1)^{n+m}{{C_nC_m}\over 2}B^{-{1\over 4}(-\alpha+2+\sqrt{1+4A})}
{{[\Gamma(1+{1\over 2}\sqrt{1+4A})]^2}\over {\Gamma(n+1+{1\over 2}\sqrt{1+4A})}}
\cr
&\times\sum\limits_{k=0}^m (-1)^k {m\choose k}{{\Gamma(k+1+{1\over 
2}\sqrt{1+4A}-{\alpha\over 2})\Gamma({\alpha\over 2}+n-k)}\over 
{\Gamma(k+1+{1\over 2}\sqrt{1+4A})\Gamma({\alpha\over 2}-k )}},
\cr\alpha < 2 + \sqrt{1+4A},\cr}\nrf
$$
with normalization coefficients $C_n$ given in Eq.(12). In case ${\alpha\over 
2}-k$ is a negative integer, then$^3$ $1/{\Gamma({\alpha\over 2}-k)}=0$ for such $k$ and the terms involving these $k$'s shall not appear in the summation of 
Eq.(27). Further, by expressing the confluent hypergeometric functions 
${}_1F_1(-n,\gamma;r)$ and ${}_1F_1(-m,\gamma;r)$ by means of the fourth formula 
in Eq.(19) and substituting these into Eq.(18), we immediately see that the sum 
appearing in Eq.(27) is a polynomial of degree $m+n$ in $\alpha$.  

With Eq.(27) we have therefore computed the matrix elements of the operator 
$x^{-\alpha}$ in the complete basis given by the Gol'dman and Krivchenkov 
eigenfunctions (9). Concomitant to our result, the matrix elements 
$$
\bra 0|x^{-\alpha}|n\ket=(-1)^nB^{\alpha\over 4}
\sqrt{{\Gamma(1+{1\over 2}\sqrt{1+4A})}\over {n!\Gamma(n+1+{1\over 
2}\sqrt{1+4A})}}{{\Gamma(1+{1\over 2}\sqrt{1+4A}-{\alpha\over 
2})\Gamma({\alpha\over 2}+n)}\over {\Gamma(1+{1\over 
2}\sqrt{1+4A})\Gamma({\alpha\over 2})}}\nrf
$$
are of special interest.
\bigskip
\noindent{\bf V. Explicit forms of the matrix elements}

\noindent In terms of the parameter $\gamma=1+{1\over 2}\sqrt{1+4A}$, the 
explicit forms of the first ten matrix elements of $x^{-\alpha}$ are for $\alpha
< 2\gamma$:
\medskip
\item\item{$x_{00}^{-\alpha}=$}$B^{\alpha\over 4}{{\Gamma(-{\alpha\over 
2}+\gamma)}\over {\Gamma(\gamma)}}$
\bigskip
\item\item{$x_{01}^{-\alpha}=$}$-{{B^{\alpha\over 4}}\over 
2}{{\Gamma(-{\alpha\over 2}+\gamma)}\over {\sqrt{\gamma}\Gamma(\gamma)}}\alpha$
\bigskip
\item\item{$x_{02}^{-\alpha}=$}${{B^{\alpha\over 4}}\over 
2^2}{{\Gamma(-{\alpha\over 2}+\gamma)}\over 
{\sqrt{2!\gamma(\gamma+1)}\Gamma(\gamma)}}\alpha(\alpha+2)$
\bigskip
\item\item{$x_{03}^{-\alpha}=$}$-{{B^{\alpha\over 4}}\over 
2^3}{{\Gamma(-{\alpha\over 2}+\gamma)}\over 
{\sqrt{3!\gamma(\gamma+1)(\gamma+2)}\Gamma(\gamma)}}\alpha(\alpha^2+6\alpha+8)$
\bigskip
\item\item{$x_{11}^{-\alpha}=$}${{B^{\alpha\over 4}}}{{\Gamma(-{\alpha\over 
2}+\gamma)}\over {4\Gamma(\gamma+1)}}(\alpha^2-2\alpha+4\gamma)$
\bigskip
\item\item{$x_{12}^{-\alpha}=$}$-{{B^{\alpha\over 4}}}{{\Gamma(-{\alpha\over 
2}+\gamma)}\over 
{8\sqrt{2(\gamma+1)}\Gamma(\gamma+1)}}\alpha(\alpha^2-2\alpha+8\gamma)$
\bigskip
\item\item{$x_{13}^{-\alpha}=$}${{B^{\alpha\over 4}}}{{\Gamma(-{\alpha\over 2}+\gamma)}\over 
{16\sqrt{3!(\gamma+2)(\gamma+1)}\Gamma(\gamma+1)}}\alpha(\alpha^3+(12\gamma-4)
\alpha+24\gamma)$
\bigskip
\item\item{$x_{22}^{-\alpha}=$}${{B^{\alpha\over 4}}}{{\Gamma(-{\alpha\over 
2}+\gamma)}\over {32\Gamma(\gamma+2)}}
(\alpha^4-4\alpha^3-16(1+2\gamma)\alpha+4(3+4\gamma)\alpha^2+32\gamma(1+\gamma))
$
\bigskip
\item\item{$x_{23}^{-\alpha}=$}${{B^{\alpha\over 4}}}{{\Gamma(-{\alpha\over 
2}+\gamma)}\over {32\sqrt{3!2!(\gamma+2)}\Gamma(\gamma+2)}}\alpha(\alpha^4-4\alpha^3+4(5+6\gamma)\alpha^2-16(2+3
\gamma)\alpha+96\gamma(1+\gamma))$
\bigskip
\item\item{$x_{33}^{-\alpha}=$}{${{B^{\alpha\over 4}}}{{\Gamma(-{\alpha\over 2}+\gamma)}\over {384\Gamma(\gamma+3)}}
(\alpha^6-8\alpha^5+(72+36\gamma)\alpha^4-(208+144\gamma)\alpha^3+(272+720\gamma+288\gamma^2)
\alpha^2-(192+1152\gamma+576\gamma^2)\alpha+384\gamma(1+\gamma)(2+\gamma)).$
}
\medskip
\noindent These matrix elements can be compared for the special case $(A, B)=(0,1)$ with the matrix elements computed by the simple harmonic oscillator 
representation supplemented by Dirichlet boundary condition [1]. We have found an error in the value of the matrix element $x_{33}^{-\alpha}$ as given by 
[1]; this error is confirmed by a re-computation according to the matrix element expression given by them. Indeed, the matrix element $x_{33}^{-\alpha}$ 
should read:
$$x_{33}^{-\alpha}={{\Gamma({{3-\alpha}\over 2})}\over {7!\Gamma({3\over 
2})}}(\alpha^6-6\alpha^5+106\alpha^4-384\alpha^3+2080\alpha^2-3408\alpha+5040)$$
instead of
$$x_{33}^{-\alpha}={{\Gamma({{3-\alpha}\over 2})}\over {7!\Gamma({3\over 
2})}}(\alpha^6-6\alpha^5+106\alpha^4-454\alpha^3+1660\alpha^2-3968\alpha+5040)$$
as quoted by Aguilera-Navarro {\it et al}$^1$.

In this work we have proved that the Gol'dman-Krivchenkov wavefunctions constitute an orthonormal basis for the Hilbert space $L^2(0,\infty)$. Using this 
orthonormal basis, we are able to construct the matrix elements of $x^{-\alpha}$: the general result (Eq.27) is convenient for use in any 
practical application which involves such singular potential terms.  It is also interesting that, with minor changes$^8$,  essentially involving only the value of the coefficient $A$, the same formulas apply immediately to the 
corresponding problems with non-zero angular momentum and in arbitrary spatial 
dimension $N \geq 2.$  A detailed variational analysis of the spiked harmonic 
Hamiltonian operator based on these matrix elements is presently in progress.  
\bigskip
\noindent{\bf Acknowledgment}
\medskip Partial financial support of this work under Grant No. GP3438 from the Natural Sciences and Engineering Research Council of Canada is gratefully 
acknowledged.
\bigskip
\noindent{\bf References}
\medskip
{\parindent=20pt

\noindent ${}^1$V. C. Aguilera-Navarro, G.A. Est\'evez and R. Guardiola. J. Math. Phys. 31, 99-104 (1990).

\noindent ${}^2$I. I. Gol'dman and D. V. Krivchenkov, {\it Problems in Quantum mechanics} (Pergamon, London, 1961).

\noindent ${}^3$L. J. Slater, {\it Confluent Hypergeometric Functions} (At the University Press, Cambridge, 1960); F. W. Sch\"afke, {\it Einf\"uhrung in die Theorie der Speziellen Funktionen der Mathematischen Physik} (Springer-Verlag, Berlin, 1963) Satz 1, p.162. 

\noindent ${}^4$L. D. Landau and M. E. Lifshitz, {\it Quantum Mechanics: Non-relativitic theory} (Pergamon, Oxford, 1977). 

\noindent ${}^5$W. Rudin, {\it Real and Complex Analysis,} $3^{rd}$ (McGraw-Hill, New York, 1987) Theorem 1.3.4, p.26; Fubini-Tonelli theorem is discussed in p.164-166.

\noindent ${}^6$D. V. Widder, {\it The Laplace Transform} (Princeton University Press, Princeton, 1972) Corollary 9.3b, p. 80; G. Doetsch, {\it Handbuch der Laplace Transformation} (Band I) (Basel, Birkh\"auser, 1971) Satz 4, p.74.

\noindent ${}^7$H. Behnke and F. Sommer,{\it Theorie der Analytischen Funktionen einer Komplexen Ver\"anderlichen} (Springer-Verlag, Berlin, 1976) Satz 28, p.138. \par}

\noindent ${}^8$R. L. Hall and N. Saad, J. Chem. Phys. {\bf 109,}
2983-2986 (1998)
\vfil\break

\hbox{\vbox{\psfig{figure=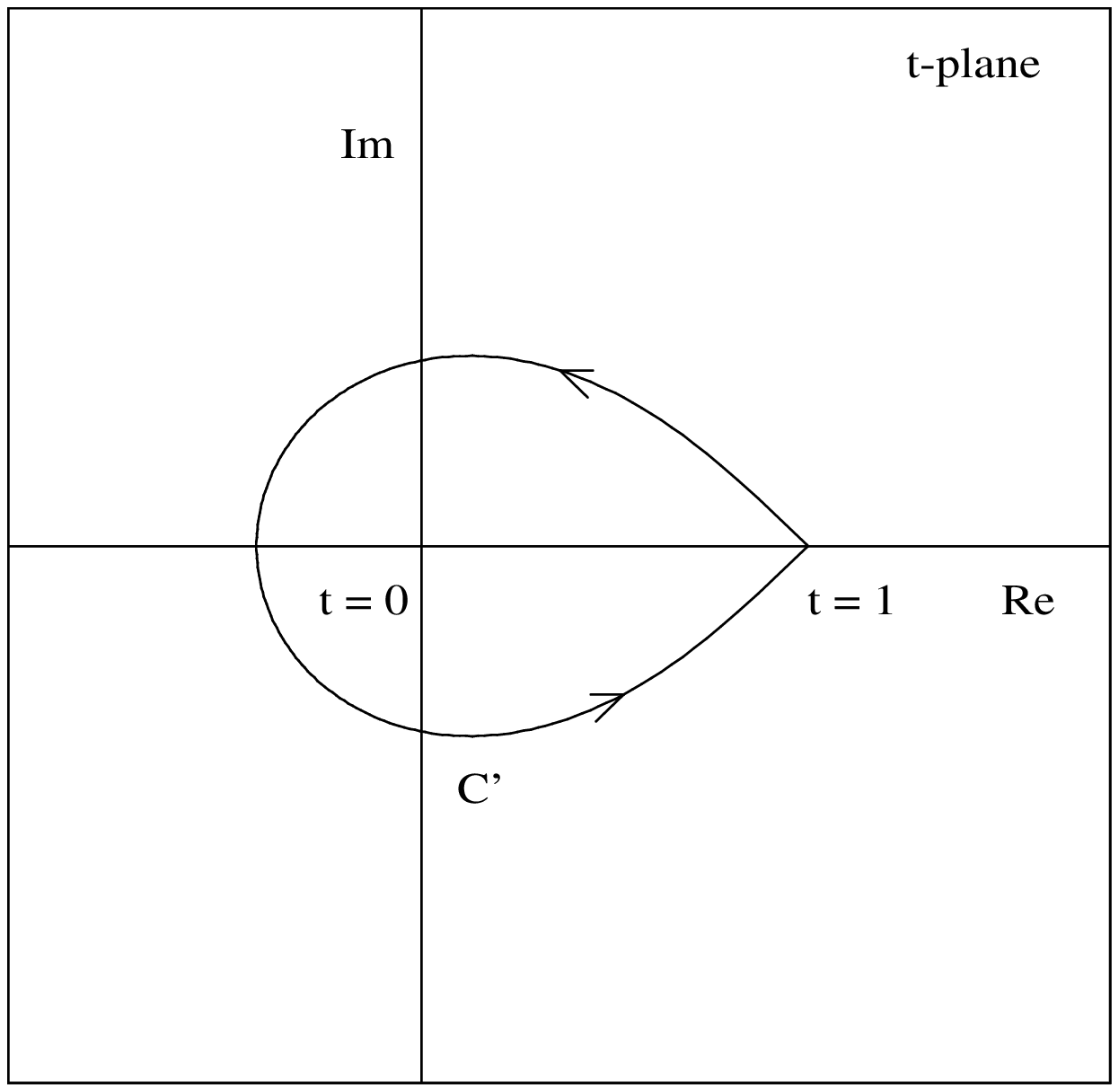,height=4in,width=4in,silent=}}}
\noindent{\bf Figure 1} The Contour $C^\prime$ in the $t$-plane starting at $1$ and enclosing the line segment from $0$ to $1$.
\vfil
\end